\documentclass[
 reprint, 
superscriptaddress, 
 amsmath,amssymb,
 aps, physrev
]{revtex4-2}

\usepackage{graphicx}
\usepackage{dcolumn}
\usepackage{bm}
\usepackage{booktabs}
\usepackage{float}

\begin{document}

\preprint{APS/123-QED}

\title{\textbf{Oxygen-nonstoichiometry-driven phase transition in Sr$_{1-x}$Nd$_{x}$CoO$_{3-\delta}$ ($x = 0.1, 0.2, 0.3$) perovskites} 
}%

\author{Nina Tereshko}
    \thanks{These authors contributed equally to this work.} 
    
  \email[\\Corresponding author: ]{nina.tereshko@gmail.com}
\author{Roman Lanovsky}
 \thanks{These authors contributed equally to  this work.}
 
 \email[\\Corresponding author: ]{ raman.lanouski@gmail.com}
 \affiliation{Institute of Solid State Physics and Semiconductors, P. Brovka street 19, 220072 Minsk, Belarus}

\author{Olivier Toulemonde}
\affiliation{CNRS, Université de Bordeaux, Bordeaux INP, ICMB UMR 5026, Pessac F-33600, France}
\author{Maxim Bushinsky}
\affiliation{Institute of Solid State Physics and Semiconductors, P. Brovka street 19, 220072 Minsk, Belarus} 
\author{Stanislav Savvin}
  \affiliation{Institut Laue-Langevin, 71 Avenue des Martyrs, 38042 Grenoble, France}
\author{Vadim Sikolenko} 
     \affiliation{Joint Institute for Nuclear Research, Dubna, 141980, Russia}
     \affiliation{Kant Baltic Federal University, 236016, Kaliningrad, Russia}
     \affiliation{Karlsruhe Institute of Technology, Karlsruhe 76131, Germany}
\author{Lingyan Xu}
     \affiliation{State Key Laboratory of Solidification Processing, Xi'an, People's Republic of China}
  \author{Aleksander Nikitin}
    \affiliation{Institute of Solid State Physics and Semiconductors, P. Brovka street 19, 220072 Minsk, Belarus}   

\date{\today}

\begin{abstract}
We report a systematic study of the interplay between oxygen nonstoichiometry, crystal structure, and magnetic and electrotransport properties in Sr$_{1-x}$Nd$_{x}$CoO$_{3-\delta}$ ($x = 0.1, 0.2, 0.3$).  High-resolution neutron powder diffraction combined with synchrotron x-ray powder diffraction reveals that increasing the oxygen content induces a structural transition from a layered $I4/mmm$ ($2a_p \times 2a_p \times 4a_p$) phase to an oxygen-deficient orthorhombic $Pmmm$ ($a_p \times a_p \times 2a_p$) phase with preferential oxygen-vacancy occupation. This transition is accompanied by a crossover from a \textit{G}-type antiferromagnetic with a weak ferromagnetic component to a ferromagnetic state, and a drastic decay in resistivity.  The evolution of the magnetic and transport properties is discussed in terms of changes in the Co spin state, enhanced Co~3\textit{d}--O~2\textit{p} orbital overlap upon oxygen uptake, and a magnetically inhomogeneous ferromagnetic state associated with residual oxygen vacancies and mixed Co$^{3+}$/Co$^{4+}$ valence.
Our findings experimentally confirm that the stabilization of the layered 314 structure is driven by the presence and ordering of oxygen vacancies rather than \textit{A}-site cation ordering, whereas the oxygen-deficient oxidized compounds represent an intermediate orthorhombic state preceding fully stoichiometric phases.

\vspace{10pt}
DOI: 10.1103/h1p6-vpb4

\end{abstract}

\maketitle

\section{Introduction}

Cobalt oxides with a perovskite-related structure can form a rich variety of compounds, both oxygen-stoichiometric and nonstoichiometric \cite{raveau2012, okimoto2021spin}. 
One of the distinctive features of cobaltites among perovskites is the relatively facile switching between the spin states of trivalent and tetravalent cobalt, originating from the strong competition between the crystal-field splitting energy and the intra-atomic (Hund) exchange interaction  \cite{korotin1996}. This competition is reflected  in phenomena such as temperature-driven spin crossover \cite{okimoto2021spin}. 

SrCoO$_{3-\delta}$ can adopt various structural configurations depending on the synthesis conditions and oxygen content. SrCoO$_{2.5}$ exists in a brownmillerite-type structure with an orthorhombic unit cell \cite{Le_Toquin_SrCoO25, Munoz_SrCoO25_PhysRevB.78.054404}. At $\delta=0.15 - 0.2$, SrCoO$_{3-\delta}$ preserves a tetragonal unit cell \cite{Nemudry_1996}, which becomes cubic for stoichiometric SrCoO$_3$ \cite{Long_2011}. The transition from cubic SrCoO$_3$ to the anion-deficient SrCoO$_{2.5}$ brownmillerite structure is accompanied by both a transition from ferromagnetic (FM) state with $T\mathrm{_C}\approx 305$~K to an antiferromagnetic (AFM) state with $T\mathrm{_N} \approx 537$~K and metal-insulator transition  \cite{Munoz_SrCoO25_PhysRevB.78.054404, Long_2011}. 

One of the interesting representatives of cobaltites with a perovskite-related structure is the so-called 314 phase with the chemical composition Sr$_3$RCo$_4$O$_{10.5}$ (Sr$_{0.75}$R$_{0.25}$CoO$_{2.625}$, where R is a rare-earth element or Y) \cite{WITHERS2003198, Istomin2003, Sheptyakov_PhysRevB.80.024409, Lanovsky_pssb.202100636, Ishiwata_PhysRevB.75.220406}. The crystal structure of these anion-deficient cobaltites is tetragonal $I4/mmm$ with a $2a_p \times 2a_p \times 4a_p$ unit cell, formed by alternating anion-deficient layers and oxygen-enriched octahedral CoO$_6$ layers touching the apexes. These compounds exhibit \textit{G}-type
long-range AFM ordering with a weak FM component \cite{JAMES20072233, Sheptyakov_PhysRevB.80.024409}.  

Pioneering studies of these cobaltites were based on the assumption that the A-site cations should be ordered \cite{Istomin2003, WITHERS2003198}. However, a series of later studies \cite{Marik_Toulemonde_oxygen, Fernandez_Toulemonde} demonstrated that the stabilization of the $I4/mmm$ structure can be promoted with only one type of cation occupying the A-site, solely through the ordering of oxygen vacancies. This important finding highlights the strong sensitivity of perovskite-related cobaltites to the concentration and ordering of oxygen vacancies.

Previously, the majority of studies focused on the Y-314 system, in which these compounds were first described. However, only a few investigations have focused on significantly modifying the oxygen content. The annealing and synthesis conditions under which oxygen contents higher than the nominal 2.625 is obtained deserve a closer look. Sheptyakov \textit{et al.}   \cite{Sheptyakov_PhysRevB.80.024409} obtained the 314 phase with a higher oxygen content by heating an as-prepared sample at 600~$^{\circ}$C for 12 h under an oxygen pressure of 150~atm. This resulted in two samples with oxygen contents of 2.63 and 2.69 for the as-prepared and oxidized batches, respectively. Somewhat higher oxygen content was reported by Raveau \textit{et al.} \cite{Raveau_2005}, achieved by sintering in an oxygen-enriched atmosphere followed by additional annealing steps. However, their work focused primarily on the magnetic and transport properties, without comparing samples with an identical cation composition but different oxygen contents. 

In the studies of  Sr$_{0.9}$Ho$_{0.1}$CoO$_{3-\delta}$ \cite{Streule_HoSrCo_PhysRevB.73.024423}, $\delta=$~0.15 was achieved by annealing at 500~K under an oxygen pressure of 800~bars for 24 h, followed by a slow cooling at a rate of 1~K/hour. More anion-deficient samples were obtained through additional reduction steps, either by annealing at 773~K under 20--50 bar of oxygen pressure or by annealing in evacuated quartz ampoules together with a getter.  For various oxygen contents, a cubic $Pm\overline{3}m$ to tetragonal $I4/mmm$ phase transition was observed, accompanied by a secondary brownmillerite $Imma$ phase in some regions of the phase diagram \cite{Streule_HoSrCo_PhysRevB.73.024423}. 

These studies qualitatively show how sensitive the 314-like cobaltites are to minor changes in the oxygen content. However, systematic studies on the influence of oxygen nonstoichiometry while maintaining an identical cation composition over a certain  substitution range remain scarce. In particular, the interplay between the oxygen-vacancy concentration, structural transformations, and magnetic properties in partially substituted systems has not yet been explored in detail. 

In the present work, we address this question by studying Sr$_{1-x}$Nd$_x$CoO$_{3-\delta}$ ($x = 0.1, 0.2, 0.3$) cobaltites prepared with different oxygen contents. The aim  of this study is twofold: (i) to perform a systematic investigation of the effect of oxygen nonstoichiometry on the crystal and magnetic structures and its interplay with magnetic and electrotransport properties; and (ii) to provide experimental confirmation that oxygen vacancies are the driving force for the stabilization of the 314-type phases without A-site cation ordering.

\section{Experimental details}

\subsection{Synthesis of the samples and their characterization}

Polycrystalline samples of Sr$_{1-x}$Nd$_{x}$CoO$_{3-\delta}$ solid solutions (\textit{x}~=~0.1,~0.2,~0.3) were prepared by the solid-state reaction method from a mixture of high-purity  Nd$_2$O$_3$, SrCO$_3$, and Co$_3$O$_4$ oxides and carbonates in the stoichiometric ratio. To remove moisture, all precursors were preannealed for 2 h prior to weighing. The powders were then mixed in a planetary ball mill (RETSCH) for 4 h at a speed of 270~rpm. 

The obtained powders were pressed into pellets and placed on platinum substrates covered with inverted corundum crucibles for the initial synthesis step, which was carried out at 1000~$^{\circ}$C for 17 h in air. Next, the pellets were crushed, reground, and pressed into two new pellets of identical size for each composition. The final synthesis was carried out in air at 1200~$^{\circ}$C for 6 h, followed by cooling to room temperature at a rate of 130~K/h. 

In order to obtain  a variant with a higher oxygen content for each composition, one of the two pellets (synthesized simultaneously from the same batch to eliminate possible variations in the synthesis procedure during the initial stages) was additionally saturated with oxygen. The oxygen saturation was performed in a tube furnace under an oxygen pressure of approximately 1.01 atm at 950~$^{\circ}$C for 36 h, followed by furnace switch-off. 

Preliminary room-temperature phase analysis was performed using x-ray powder diffractometer DRON-3M with a scanning step of 0.02$^{\circ}$ and an exposure time of 5~s (Bragg--Brentano geometry, Cu~K$\alpha$ radiation). The final confirmation of phase purity was obtained from the synchrotron and neutron diffraction experiments. 

The oxygen content of the as-prepared and oxidized samples was determined by iodometric titration. About 80--100 mg of the finely powdered sample was dissolved in a dilute hydrochloric acid solution containing KI (always kept above metallic Zn in a dark place and never exposed to air). The released elemental iodine was titrated with a standard Na$_{2}$S$_{2}$O$_{3}$ solution. The titration end point was detected using a starch indicator.

For convenience, we will refer to the samples as follows: we will use formulas reduced to the standard perovskite ABO$_3$ notation, and instead of the full chemical formula Sr$_{1-x}$Nd$_x$CoO$_{3-\delta}$ with the exact oxygen content, abbreviated labels such as N1, N2, N3, N1$_{\mathrm{oxy}}$ etc. will be used.  The number reflects the Nd content, ``oxy'' denotes the oxidized samples, and the as-prepared samples are left without ``oxy''.

\subsection{Synchrotron and neutron powder diffraction}

To confirm the phase purity and establish the crystal structure of both the as-prepared and oxidized batches of samples, synchrotron x-ray powder diffraction (SXRPD) measurements ($\lambda = 1.25$~\AA, $T = 25-300$~K) were carried out at the KMC-2 beamline of the BESSY II facility at the Helmholtz-Zentrum Berlin (HZB), Germany \cite{tobbens2016_KMC_BESSY}.  

A neutron powder diffraction (NPD) experiment for Sr$_{0.8}$Nd$_{0.2}$CoO$_{3-\delta}$ was performed using the high-resolution powder diffractometer for thermal neutrons (HRPT) at the Paul Scherrer Institute (PSI), Villigen, Switzerland \cite{FISCHER2000146_HRPT}, with an instrumental resolution of $\delta d/d < 0.001$. Measurements were carried out at temperatures of 1.5, 85, and 300~K for both as-prepared and oxidized samples using two distinct wavelengths: thermal neutrons ($\lambda = 1.494$~\AA) for the crystal structure refinement and cold neutrons ($\lambda = 2.45$~\AA) for the magnetic structure refinements. 

The diffraction data were refined using the Rietveld method \cite{rietveld1969profile} with the FullProf \cite{RODRIGUEZ_fullprof} software suite. The peak profiles were described using a standard pseudo-Voigt function for synchrotron data and the Thompson-Cox-Hastings pseudo-Voigt function \cite{NPR7func} for NPD data. Magnetic symmetry analysis was conducted using the SARA\textit{h} program \cite{Wills:gar5004}.

\subsection{Magnetic and electrotransport measurements}

Bulk magnetic measurements were performed using a commercial physical property measurement system (Cryogen Free Measurement System, Cryogenic Ltd.) in magnetic fields up to 14~T and within the temperature range of 5--300~K. Electrotransport measurements were performed using the standard four-probe method with indium contacts deposited by ultrasonic soldering.

\section{Results and discussion}

Based on the iodometric titration, the refinement of the NPD data, and the pronounced structural transformation observed upon oxidation, which will be described below, it can be concluded that the applied annealing protocol increases the oxygen content in Nd-based 314 layered cobaltites.

The oxygen indices obtained for the as-prepared and oxidized samples are summarized in Table~\ref{tab:oxygen_iodo}. For the $x=0.2$ composition, the values of the oxygen content obtained from neutron powder diffraction (NPD) refinements are in good agreement with those obtained by iodometric titration. In general, it is clear that with increasing Nd$^{3+}$ concentration, oxygen saturation becomes progressively less effective, which is associated with the difficulty of obtaining cobaltites with a concentration of Co$^{4+}$ greater than 50\% \cite{cooperGoodenough2003}. Approaching a fully oxygen-stoichiometric composition requires annealing under higher oxygen pressures or electrochemical oxidation similar to the methods described in Refs.~\cite{Bezdicka_SrCoO3_fully_st, CHENNABASAPPA20206067}.

\begin{table}[]
\caption{The oxygen stoichiometry and corresponding average charge states of Co ions (Co$^{y+}$) for the as-prepared and oxidized Sr$_{1-x}$Nd$_x$CoO$_{3-\delta}$ samples obtained via iodometry. For the N2 and N2$_\mathrm{oxy}$ samples, oxygen indices obtained from the NPD refinement are indicated in parentheses. The average error for the iodometric titration is about $\pm 0.024$.}
    \centering
    \begin{ruledtabular}
    \begin{tabular}{c | c c c c c c}
         & N1 & N2 & N3 & N1$\mathrm{_{oxy}}$ & N2$\mathrm{_{oxy}}$ & N3$\mathrm{_{oxy}}$\\ \hline
         x & 0.1 & 0.2 & 0.3 & 0.1 & 0.2 & 0.3 \\
         y & 3.28 & 3.28 & 3.40 & 3.46 & 3.46 & 3.50 \\
         O$_{3-\delta}$ & 2.69 & 2.74 (2.71) & 2.85 & 2.78 & 2.83 (2.84) & 2.90 
    \end{tabular}
    \end{ruledtabular}
    
    \label{tab:oxygen_iodo}
\end{table}

\subsection{Crystal and magnetic structures}

\textit{Crystal structure:} 
The synchrotron x-ray diffraction patterns of the as-prepared N1, N2, and N3 samples at all measured temperatures can be well indexed using a tetragonal 2$a_p \times 2a_p \times 4a_p$ unit cell within the $I4/mmm$ space group, consistent with the expected layered structure of the 314 family. Attempts to refine the patterns using alternative space groups, such as $P4/mmm$,  $A2/m$, or $Cmma$, did not yield a better fit and often led to significantly worse refinements or divergent solutions.

The basic crystal structure of the layered 314 cobaltites is formed by alternating anion-deficient and oxygen-enriched octahedral layers touching the apexes, as first independently described by Istomin \textit{et al.} \cite{Istomin2003} and Withers \textit{et al.} \cite{WITHERS2003198}.

The difference in the neutron scattering lengths of Sr ($b_{coh}=7.02$ fm) and Nd ($b_{coh}=7.69$ fm) is insufficient to reliably refine their possible site ordering from the NPD patterns. However, complementary synchrotron x-ray diffraction data allowed us to test different A-site configurations. Attempting to force a fully ordered model with Nd$^{3+}$ ions adopting the $4e$ $(0, 0, z)$ Wyckoff site resulted in  anomalously elevated values of its isotropic displacement parameter, suggesting an overestimation of electron density at that position. Additionally, the model with randomly distributed A-site ions is consistent with the findings of Refs.~\cite{Marik_Toulemonde_oxygen, Fernandez_Toulemonde}, and the present work focuses primarily on the relationship between oxygen content and related phase transitions. In this model, Nd$^{3+}$ and Sr$^{2+}$ ions statistically occupy all 3 Wyckoff sites: two $4e$ $(0, 0, z)$ and $8g$ $(0, 0,5, z)$. 

Previously it was proposed that in 314 cobaltites one of the oxygen positions should be doubled \cite{Sheptyakov_PhysRevB.80.024409, Baszczuk_HoSrCo_PhysRevB.76.134407}. During the refinement of the high-temperature NPD data, we did not observe any improvement in the quality of the fit or any other reason for doubling one of the oxygen positions. Therefore, all subsequent refinements were carried out using a structural model with four oxygen sites, similar to the one used in Refs.~\cite{Marik_Toulemonde_oxygen, Lanovsky_pssb.202100636}. The oxygen occupancy was refined for the NPD pattern collected at 300~K and fixed afterward for the remaining temperature points. It was found that the majority of oxygen vacancies are located at the O2 ($8i$) position and the minority at the O3 ($8j$) site, lying within the  $ab$ plane. 

According to Refs.~\cite{Marik_Toulemonde_oxygen, Fernandez_Toulemonde}, the stabilization of the $I4/mmm$ structure is driven by the presence and ordering of oxygen vacancies. Thus, by strongly oxidizing the samples, one can expect the destruction of the layered tetragonal $I4/mmm$ structure. Expected space-group candidates after such oxidation are a simple tetragonal $P4/mmm$ ($a_p \times a_p \times 2a_p$) or a cubic $Pm\overline{3}m$ structure, following a series of transformations proposed in Ref.\cite{Lanovsky_pssb.202100636}. Additionally, fully stoichiometric SrCoO$_3$ is known to adopt a cubic structure \cite{Long_2011, Bezdicka_SrCoO3_fully_st}.

\begin{figure}
    \centering
    \includegraphics{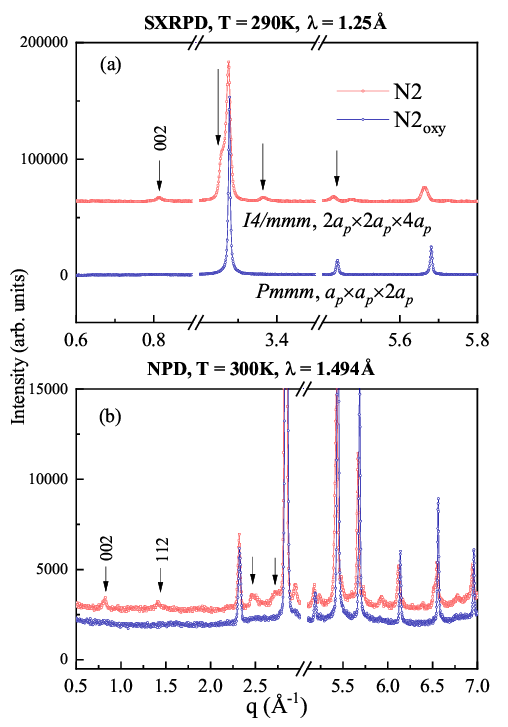}
    \caption{Selected regions of the SXRPD patterns (a) and high-resolution NPD patterns (b) of the N2 and N2$_{\mathrm{oxy}}$ samples. The data are presented in \textit{q}-space. Arrows highlight certain peaks that can be fitted only using the $I4/mmm$ space group model.}
    \label{fig:fig1_patterns_syn_neu}
\end{figure}

Figure~\ref{fig:fig1_patterns_syn_neu} shows the room-temperature SXRPD and NPD data for the representative N2 and N2$_{\mathrm{oxy}}$ samples. Indeed, it can be clearly seen that the N2$_{\mathrm{oxy}}$ sample exhibits a different crystal structure with a smaller unit cell compared to the parent N2 sample. In particular, the $002$ superstructural Bragg reflection, which is prominent for N2  and characteristic of the layered structure, is completely absent in the N2$_{\mathrm{oxy}}$ pattern.

Initial analysis in the Le Bail mode based on simple cubic or tetragonal $P4/mmm$ models yields unsatisfactory reliability factors and fit quality ($\chi^2$= 6.50 and 12.45, respectively). The $Pm\overline3m$ model failed to reproduce the asymmetric peak profiles systematically observed for several reflections of the oxidized N2$_\mathrm{oxy}$ sample, whereas the high-symmetry model predicts single unsplit reflections. Representative enlarged regions that compare the refinements within the cubic and $Pmmm$ structural models, as well as additional refinement details, are shown in  the Appendix A, Fig.~\ref{fig:appendix_fit_cube_orth}. 
A two-phase model consisting of a simple cubic $Pm\overline3m$ and orthorhombic brownmillerite phases was also tested. No signs of the secondary brownmillerite phase were found. 

The crystal structure of Sr-containing rare-earth cobaltites has previously been described implementing several space groups with a smaller unit cell compared to the tetragonal $I4/mmm$ structure. These include the monoclinic space groups $I2/a$ and $P2_1/n$ \cite{Takami_2007, Leighton_Lynn_PhysRevB.79.214420, Lanovsky_2024_PhysRevMaterials.8.114422}, the tetragonal $P4/mmm$, and several orthorhombic space groups \cite{fauth2001interplay, Plakhty_Barilo_TbBaCo_PhysRevB.71.214407, JAMES20072233, Takami_2007}. The best fit was achieved using the $Pmmm$ space group. 

As shown in Ref.~\cite{Plakhty_Barilo_TbBaCo_PhysRevB.71.214407} for TbBaCo$_2$O$_{5.5}$, two distinct phases having the same $Pmmm$ space group  but different unit cell parameters may effectively coexist: the so-called ``122'' phase with  $a_p\times2a_p\times2a_p$ unit cell and the 112 phase with $a_p\times a_p\times2a_p$ unit cell. The difference in the oxygen content between these phases is relatively small. Therefore, two possible  structural models within the $Pmmm$ space group were additionally tested. The model with the 122 cell resulted in significantly worse agreement factors than the 112 model ($\chi^2=7.40$ compared to 3.36 for the room-temperature Rietveld refinement of the NPD pattern at  $\lambda = 1.494$~\AA). Furthermore, there are no indications of any low-intensity reflections that would require an enlargement of the unit cell. Thus, for oxidized samples the patterns were described using the $Pmmm$ space group model with $a_p\times a_p\times2a_p$ unit cell.

\textit{Magnetic structure:} 
The layered structure of the 314 cobaltites provides two distinct Co sites, which allows for a variety of possible spin configurations both within each individual layer and between them. The positions of the magnetic reflections coincide with those of the nuclear reflections, indicating that the propagation vector of the magnetic structure for the tetragonal Nd-314 samples is $\mathbf{k}=[0,0,0]$.

\begin{figure*}
    \centering
    \includegraphics{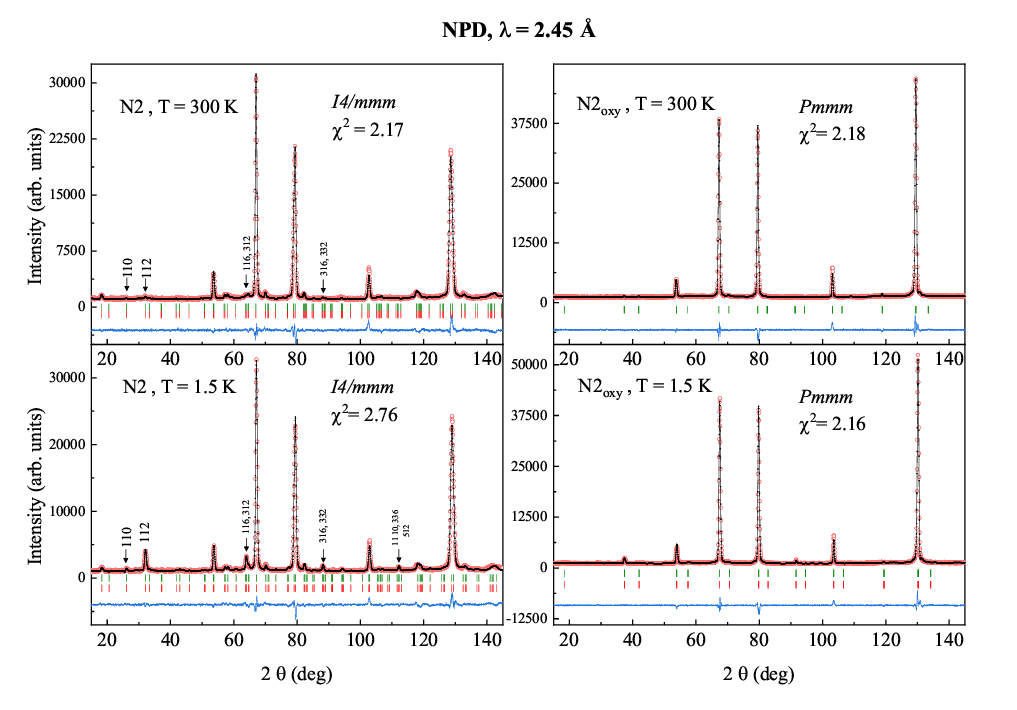}
    \caption{Observed (red circles), calculated (black line) and difference patterns of high-resolution NPD data for the as-prepared (N2, left panels) and oxidized (N2$_{\mathrm{oxy}}$, right panels) Sr$_{0.8}$Nd$_{0.2}$CoO$_{3-\delta}$ samples, collected at $\lambda = 2.45$~\AA. The vertical bars indicate the calculated positions of the nuclear (olive) and magnetic (red) Bragg peaks ($\mathbf{k}=[0,0,0]$). For the N2, the indices of reflexes with the greatest magnetic contribution are indicated.}
    \label{fig:fig_NPD_245_magn}
\end{figure*}

\begin{table}[t]
\caption{\label{tab:mag_basis_i4} Basis functions ($\psi_n$) of the $\Gamma_5$ irreducible representation for the Co1 and Co2 crystallographic sites in the $I4/mmm$ space group and $\mathbf{k}=[0,0,0]$.}
\begin{ruledtabular}
\begin{tabular}{l rrr rrr rrr}
 & \multicolumn{3}{c}{Co1: $\psi_1$} & \multicolumn{3}{c}{Co2: $\psi_1$} & \multicolumn{3}{c}{Co2: $\psi_2$} \\
\cmidrule(lr){2-4} \cmidrule(lr){5-7} \cmidrule(lr){8-10}
Atom & $m_a$ & $m_b$ & $m_c$ & $m_a$ & $m_b$ & $m_c$ & $m_a$ & $m_b$ & $m_c$ \\
\midrule
1 & 0 & 0 &  1 &  1 &  1 &  0 & 0 & 0 &  1 \\
2 & 0 & 0 & -1 &  1 & -1 &  0 & 0 & 0 & -1 \\
3 & 0 & 0 & -1 & -1 &  1 &  0 & 0 & 0 & -1 \\
4 & 0 & 0 &  1 & -1 & -1 &  0 & 0 & 0 &  1 \\
\end{tabular}
\end{ruledtabular}
\end{table}

According to the symmetry analysis of the allowed magnetic representations for the propagation vector $\mathbf{k}=[0, 0, 0]$ (performed using the SARA\textit{h} program \cite{Wills:gar5004}), nonzero decompositions of the magnetic representations for two Co sites yield eight irreducible representations (IRs) for the Co1 site and  five IRs for the Co2 site. The best fit was achieved in the model with the magnetic moments of Co1 aligned along the \textit{c} axis with a single basis vector (BV), while the magnetic moments of Co2 have two BVs, one of them constrained within the \textit{ab} plane and the other pointing along the \textit{c} axis. The magnetic moments in both layers form a \textit{G}-type AFM order (Table~\ref{tab:mag_basis_i4}). For the Co2 site the component along \textit{c} direction is dominant, whereas the second component lying in the \textit{ab} plane is significantly smaller.

The distribution of magnetic moments between the Co1 and Co2 sites was refined taking into account physical constraints. The bond distance analysis from the pure ionic model is not so straightforward, but to some extent it can be used as a good approximation.  The average Co2--O bond length in the octahedral layer is consistent with the high-spin (HS) Co$^{3+}$, while the Co1 anion-deficient layer with a shorter Co1--O bond length  corresponds to the mixed valence and spin states, as discussed further below.  Thus, despite the fact that from the agreement factors for the models with a larger magnetic moment in both Co1 and Co2 sites are equivalent, the model with the larger magnetic moment localized at the Co2 site is preferred and is consistent with the interatomic distances in the octahedral layer and previously reported results for the Y-314 cobaltites \cite{Lanovsky_pssb.202100636}.

It should be noted that for the N2 sample, reflections $110$, $112$ and some others preserve their magnetic contribution to the NPD pattern at 300~K (see Figs.~\ref{fig:fig1_patterns_syn_neu},~\ref{fig:fig_NPD_245_magn}). Special attention should be paid to the $110$ reflection, which arises directly from the difference in magnetic moments at the two cobalt sites. It can be easily verified by artificially setting the same magnetic moment for the two cobalt positions, and $110$ reflection can be considered purely magnetic. Therefore, room temperature NPD pattern refinement was performed using a two phase (nuclear + magnetic) model as for the rest of the measured temperatures. 

\begin{table}[t]
\caption{\label{tab:mag_basis_pmmm} Basis functions $\psi_n$ of the irreducible representations $\Gamma_3$, $\Gamma_5$, and $\Gamma_7$ for the Co site in the $Pmmm$ space group with $\mathbf{k}=[0,0,0]$. All three representations correspond to a simple ferromagnetic configuration with the magnetic vector directed along the \textit{a}, \textit{b}, and \textit{c} axes, respectively. The $\Gamma_7$ model (highlighted in bold) was selected for the final refinement.}
\begin{ruledtabular}
\begin{tabular}{l rrr rrr rrr}
 & \multicolumn{3}{c}{$\Gamma_3: \psi_1$} & \multicolumn{3}{c}{$\Gamma_5: \psi_1$} & \multicolumn{3}{c}{\textbf{$\Gamma_7: \psi_1$}} \\
\cmidrule(lr){2-4} \cmidrule(lr){5-7} \cmidrule(lr){8-10}
Atom & $m_a$ & $m_b$ & $m_c$ & $m_a$ & $m_b$ & $m_c$ & $\mathbf{m_a}$ & $\mathbf{m_b}$ & $\mathbf{m_c}$ \\
\midrule
1 & 1 & 0 & 0 & 0 & 1 & 0 & \textbf{0} & \textbf{0} & \textbf{1} \\
2 & 1 & 0 & 0 & 0 & 1 & 0 & \textbf{0} & \textbf{0} & \textbf{1} \\
\end{tabular}
\end{ruledtabular}
\end{table}

\begin{table}[t]

\caption{\label{tab:magmomNPD}Temperature dependence of the refined magnetic moments (in $\mu_B$/Co) of the N2 and N2$_\mathrm{oxy}$ samples. For the $I4/mmm$ \textit{G}-type AFM model, the Co1 occupies the Wyckoff site 8\textit{h} (\textit{x, x}, 0) and Co2 the Wyckoff site 8\textit{f} (0.25, 0.25, 0.25). For the $Pmmm$ model, the FM ordering occurs with Co at the Wyckoff site 2\textit{q} ({0, 0, \textit{z}}). The propagation vector for both models is $\mathbf{k}=[0, 0, 0]$.}
\begin{ruledtabular}
\begin{tabular}{l ccc}
 & \multicolumn{2}{c}{$I4/mmm$ (\textit{G}-type AFM)} & $Pmmm$ (FM) \\
 \cmidrule(lr){2-3} \cmidrule(lr){4-4}
 Temperature & $M_{\mathrm{Co1}}$ ($\mu_B$) & $M_{\mathrm{Co2}}$ ($\mu_B$) & $M_{\mathrm{Co}}$ ($\mu_B$) \\
\midrule
1.5 K  &  1.066(05) & 2.327(05) & 1.325(12) \\
85 K  & 1.049(23) &  2.052(06) & 1.095(25) \\
300 K & 0.194(43) & 0.728(94)  & -- \\
\end{tabular}
\end{ruledtabular}
\end{table}

To better describe the magnetic structure and obtain the most reliable magnetic moment values, the diffraction patterns collected using a longer wavelength ($\lambda = 2.45$~\AA) were used for magnetic refinement, while the crystal structure parameters were fixed to those obtained from the $\lambda = 1.494$~\AA~data.  Refined NPD patterns for N2 and N2$_{\mathrm{oxy}}$ samples at 1.5 and 300~K are shown in Fig.~\ref{fig:fig_NPD_245_magn}.

For the oxidized N2$_{\mathrm{oxy}}$ sample there is only one Co site within the $Pmmm$ $a_p\times a_p\times 2a_p$ model, and the same propagation vector $\mathbf{k}=[0,0,0]$ matches. Magnetic representation analysis yields six nonzero components of decomposition of the magnetic representation. Only three candidate IRs ($\Gamma_3$, $\Gamma_5$ and $\Gamma_7$) allow to describe the observed intensities of magnetic peaks. These IRs represent a simple FM order with magnetic moments aligned along the $a$, $b$, and \textit{c} axes, respectively. 

For polycrystalline samples these models provide identical fits to the diffraction data. Therefore, the $\Gamma_7$ representation was chosen for the refinement, assuming the FM component remains parallel to the \textit{c} axis, consistent with the magnetic anisotropy observed in the tetragonal parent phase and for convenience of comparison (Table~\ref{tab:mag_basis_pmmm}). No long-range AFM component was observed in the diffraction patterns.

It is important to emphasize that oxidation of the samples leads not only to a drastic structural transition but also to a simultaneous change in the type of magnetic ordering from the \textit{G}-type AFM to a collinear FM. No long-range magnetic ordering of Nd$^{3+}$ ions was found in either the oxidized or the as-prepared samples. The refined magnetic moment values are summarized in Table~\ref{tab:magmomNPD}.

\textit{Possible reasons for symmetry lowering upon the oxidation:} One remaining question concerning the structural analysis is what drives the symmetry lowering from $I4/mmm$ to $Pmmm$ upon the oxidation and why the resulting crystal structure remains inconsistent with cubic symmetry. Although the microscopic origin of the orthorhombic distortion cannot be uniquely established from the present diffraction data, there are some points to consider.

First, our case differs significantly from the one described in Refs.\cite{CHENNABASAPPA20206067, Bezdicka_SrCoO3_fully_st}, where the La$_{1-x}$Sr$_x$CoO$_3$ and SrCoO$_3$ were electrochemically oxidized to a nearly fully stoichiometric state. As determined by the iodometric titration route and refined from NPD patterns, the oxygen content of the oxidized N$_\mathrm{oxy}$ series remains far from stoichiometry.  In addition, the oxygen vacancies are not fully disordered. The NPD refinement results of N2$_\mathrm{oxy}$ show that the oxygen vacancies are located  at only one of the four oxygen sites and the other three sites are fully occupied.  In other words, partial oxidation effectively destroys the long-range ordering of oxygen vacancies across the layers of $I4/mmm$ $2a_p\times2a_p\times4a_p$ structure. However, remaining oxygen vacancies are abundant and not randomized (see Table~\ref{tab:tab1_N2_npd}). 

Secondly, the ionic radius of Nd$^{3+}$ is considerably smaller than the ionic radius of Sr$^{2+}$ (1.27 and 1.44~\AA for CN=12, respectively). This reduces the tolerance factor and favors octahedral tilting.  

Therefore, the considered \textit{oxygen-deficient} system represents a structural state in which the long-range ordering of vacancies across layers characteristic of the 314 $I4/mmm$ cobaltites has already disappeared, whereas complete oxygen stoichiometry has not yet been achieved. 

Residual oxygen vacancies and octahedral tilting induced by Nd$^{3+}$ substitution  contribute to the stabilization of the orthorhombic structure, making it energetically more favorable than the ideal cubic perovskite which may be further stabilized by local distortions associated with mixed Co$^{3+}$/Co$^{4+}$ valence and spin states.

The values obtained for Co--O bond lengths and Co--O--Co bond lengths for N2 and N2$\mathrm{oxy}$, fruitful for discussion of the  magnetic and electrotransport properties below, are summarized in Table~\ref{tab:tab_Co_O_lengths}. Additional details regarding crystal and magnetic structure refinement, as well as the main structural parameters and residual factors for both as-prepared and oxidized samples are presented in the Appendix A.

\begin{table*}
    
    \caption{\label{tab:tab_Co_O_lengths}Co--O bond lengths and Co--O--Co bond angles obtained from the Rietveld refinement of the high-resolution NPD data  for Sr$_{0.8}$Nd$_{0.2}$CoO$_{2.71}$ and Sr$_{0.8}$Nd$_{0.2}$CoO$_{2.84}$  in the $I4/mmm$ and $Pmmm$ structural models, respectively; $\lambda=1.494$~\AA.}

    \begin{ruledtabular}
    \begin{tabular}{l  c c c l c c c }
        \multicolumn{4}{c}{$I4/mmm$, Sr$_{0.8}$Nd$_{0.2}$CoO$_{2.71}$} & \multicolumn{4}{c}{$Pmmm$, Sr$_{0.8}$Nd$_{0.2}$CoO$_{2.84}$} \\
       $T$ & 1.5~K & 85~K &  300~K & $T$ & 1.5~K & 85~K &  300~K \\ 
       Co1--O2 (\AA) & 1.913(14) & 1.917(16) & 1.904(13) &  Co--O1 (\AA) & 1.925(15) & 1.939(13) & 1.943(15) \\
       Co1--O3 (\AA) & 1.940(14) & 1.937(16) & 1.947(14) &  Co--O2 (\AA) & 1.896(15) & 1.883(13) & 1.889(15) \\
       Co1--O1 (\AA) & 1.834(5) & 1.825(5) & 1.827(5) &  Co--O3 (\AA) & 1.9086(3) & 1.9090(3) & 1.9166(4) \\
       Co1--O2--Co1 (deg) & 177.5(13) & 179.1(15) & 176.8(13) &  Co--O4 (\AA) & 1.9112(4) & 1.9110(3) & 1.9147(5) \\
       Co1--O3--Co1 (deg) & 163.8(13) & 163.7(15) & 167.0(12) &  Co--O3--Co (deg) & 178.347(10) & 178.348(10) & 177.388(16) \\
       Co1--O1--Co2 (deg) & 176.7(2) & 176.2(2) & 174.7(2) &  Co--O4--Co (deg) & 177.434(16) & 177.708(13) & 176.789(19) \\
       Co2--O1 (\AA) & 2.019(5) & 2.028(5) & 2.035(5) & Co--O1--Co (deg)  & \multicolumn{3}{c}{180} \\
       Co2--O4 (\AA) & 1.918(7) & 1.918(6) & 1.919(6) & Co--O2--Co (deg)  & \multicolumn{3}{c}{180} \\
       Co2--O4--Co2 (deg) & 175.997(14) & 176.304(13) & 177.552(9) &  &  &  &  \\
       $\langle \mathrm{Co1-O} \rangle$ (\AA) & 1.895 & 1.893 & 1.893 & $\langle \mathrm{Co-O} \rangle$ (\AA) & 1.910 & 1.911 & 1.916 \\
       $\langle \mathrm{Co2-O} \rangle$ (\AA) & 1.969 & 1.974 & 1.977 &   &  &  &  \\

    \end{tabular}      
    \end{ruledtabular}
    
\end{table*}

\subsection{Magnetic properties}

The description of magnetic interactions in Sr$_{1-x}$Nd$_x$CoO$_{3-\delta}$ requires taking into account the valence and spin states of cobalt ions.

While describing cobaltites in terms of superexchange interactions \cite{Munoz_SrCoO25_PhysRevB.78.054404, cooperGoodenough2003, gooden_ruls}, the HS Co$^{3+}$ ions interact strongly via AFM, Co$^{3+}$ ions in the intermediate-spin (IS) state interact via FM \cite{Phelan2006IS, okimoto2021spin, Lanovsky_2024_PhysRevMaterials.8.114422}. Low-spin (LS) Co$^{3+}$ ions are comparatively magnetically inactive. Additionally, both the as-prepared and oxidized samples contain a significant amount of Co$^{4+}$ ions (see Table~\ref{tab:oxygen_iodo}), promoting the FM double-exchange interaction between HS Co$^{3+}$ and Co$^{4+}$.

\begin{table*}
    \caption{\label{tab:tab1_N2_npd} Refined crystal structure parameters and agreement factors for Sr$_{0.8}$Nd$_{0.2}$CoO$_{2.71}$ (N2) and Sr$_{0.8}$Nd$_{0.2}$CoO$_{2.84}$ (N2$_{\mathrm{oxy}}$) obtained from the refinement of the high-resolution NPD data (thermal neutrons, $\lambda=1.494$~\AA) using the structural models $I4/mmm$ with a $2a_p \times 2a_p \times 4a_p$ unit cell and the $Pmmm$ with an $a_p \times a_p \times 2a_p$ unit cell, respectively. The occupancy of the A-site is proportional to the given chemical formula (model with absence of the Nd$^{3+}$/Sr$^{2+}$ ordering).}

    \begin{ruledtabular}
    \begin{tabular}{l  c c c l  c c c  }
    \multicolumn{4}{c}{$I4/mmm$, Sr$_{0.8}$Nd$_{0.2}$CoO$_{2.71}$} & \multicolumn{4}{c}{$Pmmm$, Sr$_{0.8}$Nd$_{0.2}$CoO$_{2.84}$} \\
       $T$ & 1.5~K & 85~K &  300~K & $T$ & 1.5~K & 85~K &  300~K  \\ 
        \textit{a = b}~(\AA) & 7.66697(8) & 7.66765(9)  & 7.67505(9) & \textit{a}~(\AA) & 3.82136(31) & 3.82127(32)  & 3.82782(9)   \\
        \textit{c}~(\AA) & 15.40306(28) & 15.40504(32) &  15.42936(27) & \textit{b}~(\AA) & 3.81682(8) & 3.81757(10)  & 3.83224(30)    \\
        \textit{V}~(\AA$^3$) & 905.429(21) & 905.707(25) &   908.889(22) & \textit{c}~(\AA) & 7.64199(65) & 7.64458(55) &  7.66359(63)    \\
    
       Nd1/Sr1  & \multicolumn{3}{c}{$4e$ (0,0,\textit{z})} & \textit{V}~(\AA$^3$) & 111.462(13) & 111.519(13) &   112.418(13)     \\
        \textit{z}  &  0.13581(75)  &  0.13590(72)  & 0.13450(68) & Nd1/Sr1  & \multicolumn{3}{c}{$1h$ (0.5,0.5,0.5)}   \\
       Nd2/Sr2 &  \multicolumn{3}{c}{$4e$  (0,0,\textit{z})}  & Nd2/Sr2 &  \multicolumn{3}{c}{$1f$  (0.5,0.5,0)} \\
       \textit{z}  & 0.62513(74) & 0.62510(72) & 0.62481(69) & Co & \multicolumn{3}{c}{$2q$ (0,0,\textit{z})}   \\
       Nd3/Sr3 & \multicolumn{3}{c}{$8g$ (0,0.5,\textit{z})} & \textit{z}  & 0.25191(182) & 0.25375(169) & 0.25346(188) \\
       \textit{z}  & 0.13179(51) & 0.13175(46) & 0.13287(39) & O1 & \multicolumn{3}{c}{$1a$ (0,0,0)}   \\
       Co1 & \multicolumn{3}{c}{$8h$ (x,x,\textit{0})} &  Occ & \multicolumn{3}{c}{1.0}  \\
       \textit{x}  & 0.24950(179) & 0.24998(191) & 0.24796(165) & O2 & \multicolumn{3}{c}{$1c$ (0,0,0.5)}   \\
       Co2 & \multicolumn{3}{c}{$8f$ (0.25,0.25,0.25)} & Occ & \multicolumn{3}{c}{0.674(16)} \\
       O1 & \multicolumn{3}{c}{$16m$ (\textit{x,x,z})} & O3 &  \multicolumn{3}{c}{$2r$ (0,0.5,\textit{z})} \\
       \textit{x} & 0.24459(107) & 0.24406(101) & 0.24065(98) & \textit{z} & 0.24834(77) & 0.25733(104) & 0.25924(97) \\
       \textit{z} & 0.11899(27) & 0.11838(26) & 0.11826(25) & Occ & \multicolumn{3}{c}{1.0}   \\
       Occ & \multicolumn{3}{c}{1.0} & O4 & \multicolumn{3}{c}{$2s$ (0.5,0,\textit{z})}    \\
       O2 & \multicolumn{3}{c}{$8i$ (\textit{x},0,0)} & \textit{z}  & 0.25752(103) & 0.24867(81) &  0.24653(76)  \\
       \textit{x} & 0.24440(256) & 0.24842(194) & 0.25470(239) & Occ & \multicolumn{3}{c}{1.0}\\
       Occ & \multicolumn{3}{c}{0.4640(12)}  & $\chi^2$  & 3.64 & 3.75 & 3.36 \\
       O3 & \multicolumn{3}{c}{$8j$ (\textit{x},0.5,0)}& $R_{Bragg}$ & 4.86  & 4.82 & 5.73  \\
       \textit{x} & 0.78614(142) & 0.78590(141) & 0.78080(151)  & & & & \\
       Occ & \multicolumn{3}{c}{0.9600(16)}  & & & & \\
       O4 & \multicolumn{3}{c}{$16n$ (0,\textit{y,z})} & & & &\\
       \textit{y}  & 0.25125(77) & 0.25070(69) & 0.25110(64) & & & &  \\
        \textit{z}  & 0.25430(55) & 0.25397(56) &  0.25262(54) & & & &  \\
        Occ & \multicolumn{3}{c}{1.0} & & & &  \\
       $\chi^2$  & 4.59 & 4.13 & 3.05  & & & & \\
       $R_{Bragg}$ & 7.89 & 7.47 & 7.59   & & & & \\

    \end{tabular}      
    \end{ruledtabular}

\end{table*}

A deviation of the bond angles Co--O--Co from 180$^\circ$ leads to a reduction of orbital overlap between the Co~3\textit{d} and O$^{2-}$~2\textit{p} orbitals, weakening the exchange interaction.
Thus, it is essential to focus simultaneously on orbital occupancy, exchange interactions, magnetic phase competition, and structural distortions.

\begin{figure*}
    \centering
    \includegraphics{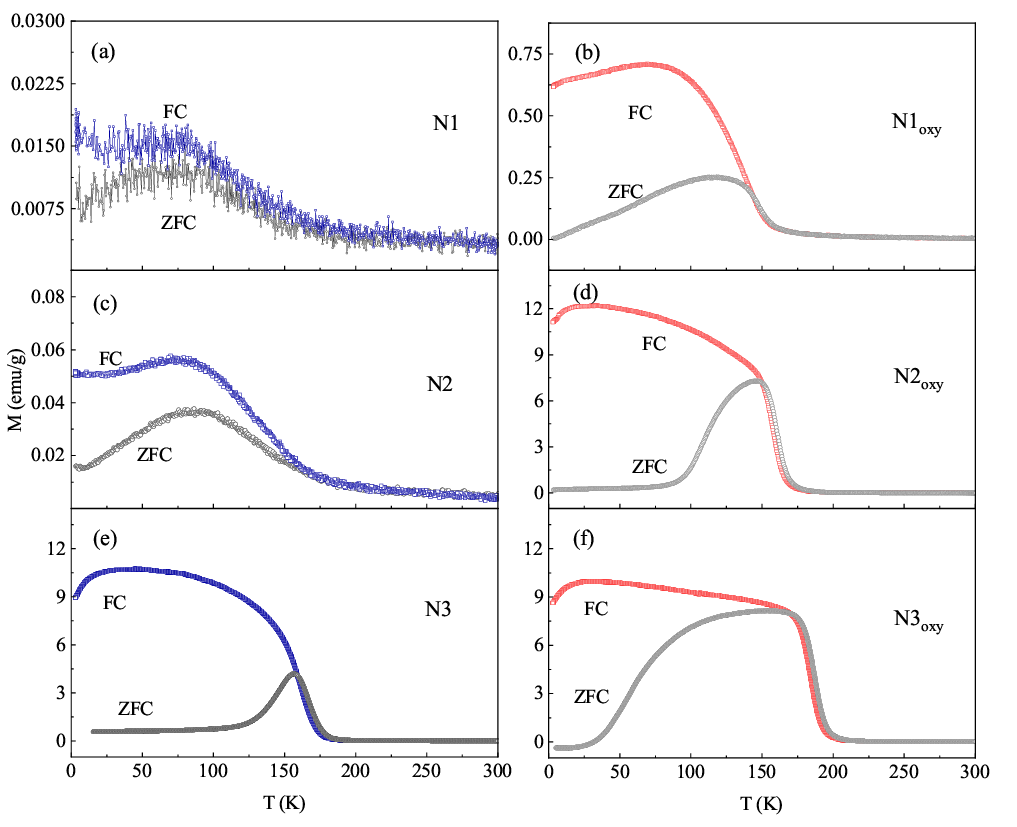}
    \caption{Temperature dependences of the magnetization measured under field-cooled (FC) and zero-field-cooled (ZFC) conditions for the as-prepared (left panels) and oxidized (right panels) Sr$_{1-x}$Nd$_x$CoO$_{3-\delta}$ (\textit{x}= 0.1, 0.2, 0.3) samples measured in an applied field of 100~Oe.}
    \label{fig:fig_FC_ZFC}
\end{figure*}

The temperature dependences of  magnetization for the as-prepared and oxidized Sr$_{1-x}$Nd$_x$CoO$_{3-\delta}$ (\textit{x}= 0.1, 0.2, 0.3) samples measured in a static magnetic field of 100~Oe are shown in Fig.~\ref{fig:fig_FC_ZFC}. The most striking feature is the orders of magnitude increase in magnetization upon oxidation. 

The as-prepared N1 and N2 exhibit very low magnetization with a divergence between the ZFC and FC curves. This is a common behavior for the layered 314 cobaltites, characterized by a \textit{G}-type AFM structure with a small FM component. Although the FM component is small, a temperature of 300~K is insufficient to suppress it completely, which is consistent with NPD data for N2 with the Néel temperature above 300~K. Co--O--Co  pathways are partially disrupted by the high concentration of oxygen vacancies, leading to magnetically frustrated spin-glass like states. 

All oxidized samples show a sharp magnetization rise, with  $T_\mathrm{C}$ increasing with Nd content, shifting from $T_\mathrm{C}\approx$~160~K for N1$_\mathrm{oxy}$ to $T_\mathrm{C}\approx$~200~K for N3$_\mathrm{oxy}$. According to iodometry, all the oxidized samples contain more than 40\% Co$^{4+}$. Thus, oxidation not only restores Co--O--Co pathways, but also triggers FM double-exchange, although magnetization values of the N1$_\mathrm{oxy}$ are an order of magnitude smaller than those of N2$_\mathrm{oxy}$ and N3$_\mathrm{oxy}$. Increasing Nd$^{3+}$ content applies chemical pressure and simultaneously facilitates oxygen uptake. The combined effect leads to changes in the Co-O bond lengths and Co--O--Co bond angles. In particular, the reduced oxygen deficiency improves the integrity of the Co--O network and enhances Co~3\textit{d}--O~2\textit{p} orbital overlap.

As an anomaly, the as-prepared N3 behaves like an FM even before formal oxidation. Its oxygen content (see Table~\ref{tab:oxygen_iodo}) is slightly below the threshold of the average oxidation state, separating weak ferromagnets with  semiconductor-like conductivity from metallic ferromagnets, proposed by Raveau \textit{et al.}\cite{Raveau_2005}. 

For the oxidized and N3 samples, a magnetization drop below $T\approx 25$~K is observed. This temperature is significantly higher than the magnetic ordering temperature of Nd$^{3+}$ ions. Nd–-Co antiferromagnetic 4\textit{f}-3\textit{d} exchange may contribute to the low-temperature decrease in magnetization, however, this cannot be the primary factor governing the magnetic evolution across the series. The NPD data revealed no signs of Nd$^{3+}$ ordering even at $T = 1.5$~K for both series and the NPD spectra are adequately described using magnetic moments on the Co sites. Moreover, similar composition-dependent magnetization behavior is observed in related cobaltites that do not contain any magnetic A-site ions \cite{Lanovsky_pssb.202100636, Marik_Toulemonde_oxygen, Raveau_2005}. These comparisons indicate that the magnetic properties correlate primarily with oxygen content, oxygen-vacancy ordering, Co valence and spin-state distribution, and Co–O–Co geometry rather than A-site magnetism itself.

All oxidized and N3 samples exhibit a certain degree of FC/ZFC irreversibility. This behavior most likely reflects magnetic inhomogeneity within the FM state associated with the presence of residual oxygen vacancies and the random distribution of mixed Co$^{3+}$/Co$^{4+}$ ions, resulting in spatial variations of the exchange interactions and domain-wall pinning. Such behavior is consistent with the incomplete magnetic saturation observed in the $M(H)$ measurements.

\begin{figure}
    \centering
    \includegraphics{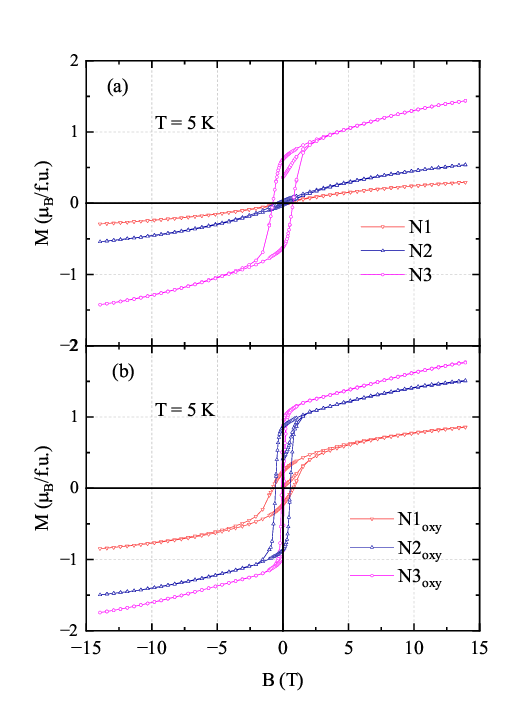}
    \caption{Field dependences of the magnetization for the as-prepared (a) and oxidized (b) Sr$_{1-x}$Nd$_x$CoO$_{3-\delta}$ (\textit{x} = 0.1, 0.2, 0.3) samples measured at $T = 5$~K.}
    \label{fig:fig_M_H_5K}
\end{figure}

The field dependences of the magnetization for the as-prepared and oxidized Sr$_{1-x}$Nd$_x$CoO$_{3-\delta}$ (\textit{x}= 0.1, 0.2, 0.3) samples measured at $T=5$~K are shown in Fig.~\ref{fig:fig_M_H_5K}.

Complete magnetic saturation of the samples is not achieved even in fields up to $B=14$~T. The N1 and N2 show an almost linear $M(H)$ dependence with a remanence of about 0.27 and 0.8~emu/g, respectively. N3 exhibits much higher absolute magnetization values, as well as a remanence of 16.5~emu/g. The field dependences of the magnetization of oxidized samples confirm the conclusions drawn from the $M(T)$ data. The reduction of coercivity $H_\mathrm{C}$ from 0.8~T for N1$_\mathrm{oxy}$ to 0.1~T for N3$_\mathrm{oxy}$ suggests that Nd$^{3+}$-doping and consequently higher oxygen content facilitates the crossover from cluster-glass-like with pinned FM clusters to a more homogeneous FM state. Values of spontaneous magnetization (in the absence of saturation) determined by tangent method are in good agreement with data refined from NPD. Let us further return to the structural data obtained from NPD to better analyze the available magnetization data.

In the layered $I4/mmm$ structure, the anion-deficient layers and the octahedral layers are connected via the apical oxygen O1 with a canted Co1--O1--Co2 angle ($\approx 175^\circ$). The octahedra of the oxygen-enriched layer are characterized by fairly straight bond angles in the $ab$ plane and are elongated along the \textit{c} axis; the average Co2--O bond lengths are about 1.97~\AA (see Table~\ref{tab:tab_Co_O_lengths}). This corresponds to a model with  HS Co$^{3+}$ ions within octahedral layer, having strong AFM superexchange interactions. The presence of abnormally short Co1--O1 bond lengths within the anion-deficient layers, as well as the average Co1--O bond length of about 1.893~\AA, indicates the localization of Co$^{4+}$ in this layer. Concurrently, Co1--O3--Co1 bond angle is strongly canted, which worsens the Co~3\textit{d}--O~2\textit{p} orbital overlap and weakens the potential strength of FM Co$^{4+}$--O$^{2-}$--Co$^{3+}$ double-exchange, while the second oxygen position O2 within the $ab$ plane is half filled.  This leads to the presence of weak short-range ferromagnetism within the anion-deficient layers, while octahedral layers exhibit a strong AFM nature. In the magnetic dependences, this is reflected as a small ferromagnetic component with a spin-glass-like $M(T)$ dependence.

A significantly different situation is observed for the oxidized N2$_\mathrm{oxy}$. Two of the Co--O--Co bond angles are exactly 180$^\circ$, while the other two are only insignificantly distorted. There is only one Co site with average Co--O bond length of about 1.91~\AA. In the presence of $\sim 46$\% Co$^{4+}$ ions according to the iodometry and NPD data refinement, similar values of the average ionic radius can be realized via the following two scenarios: \\
(i) IS~Co$^{3+}$~+~Co$^{4+}$ with approximate chemical formula Sr$_{0.8}$Nd$_{0.2}$(IS)Co$^{3+}_{0.54}$Co$^{4+}_{0.46}$O$_{2.84}$; \\ (ii) and mixed spin-state scenario LS/HS~Co$^{3+}$~+~Co$^{4+}$ with composition Sr$_{0.8}$Nd$_{0.2}$(LS)Co$^{3+}_{0.21}$(HS)Co$^{3+}_{0.33}$Co$^{4+}_{0.46}$O$_{2.84}$.

Scenario (i) is expected to have strong ferromagnetism and metallic-type conductivity with a broad $e_g$ band, whereas scenario (ii) is more complex. HS Co$^{3+}$ -- Co$^{4+}$ interactions favor the double-exchange mechanism, while the LS Co$^{3+}$ admixture acts as a diamagnetic insulator and reduces the connectivity of the HS Co$^{3+}$--Co$^{4+}$ pathways. The exact spin state of Co$^{4+}$ (LS or IS, or some mixed spin state) in these oxygen-deficient layered cobaltites cannot be unambiguously determined. Nevertheless, the proposed interpretation of the magnetic evolution does not depend critically on the specific spin-state assignment of Co$^{4+}$.
Taking into account the presence of cluster glass behavior in the $M(T)$ curves, as well as the nature of the $M(H)$ dependences, the most probable model is scenario (ii), in which the Co$^{3+}$ ions are in a mixed LS/HS state.

In addition, another possibility for describing the properties of highly oxidized samples must be considered. In Ref.~\cite{CHENNABASAPPA20206067}, during the consideration of a nearly fully stoichiometric La$_{1-x}$Sr$_x$CoO$_3$, an itinerant electron model \cite{Bezdicka_SrCoO3_fully_st} was proposed to describe the observed properties. Their electrochemically oxidized samples exhibit FM behavior with a relatively high magnetic moment per Co ion and $T_\mathrm{C}$ up to 280~K. This value is higher than the maximum $T_\mathrm{C}$~=~250~K suggested by Wu and Leighton \cite{Wu_Leighton_PhysRevB.67.174408} for IS Co$^{3+}$ to IS Co$^{4+}$ 1:1 ratio in a pure double-exchange model. In our case, the Curie temperature of the oxidized samples is lower, as are the values of the magnetic moment per Co ion. Applying the itinerant electron model proposed by Bezdicka~\textit{et~al.}~\cite{Bezdicka_SrCoO3_fully_st} with $M\mathrm{_{sat}}$~=~(1+2x)$\mu_B$, this would correspond to a fraction of delocalized electrons of about 0.16 for N2$_\mathrm{oxy}$, and a small portion of the Co$^{4+}$ ions could be stabilized in the IS state. Although this value is fairly low, there is no reason to completely exclude this contribution. A complex mixed model, in which superexchange interactions and double exchange dominate is possible, but a contribution according to the itinerant electron model is also present.
 
Since a LS Co$^{3+}$ fraction of 21\% should be insufficient to disrupt the percolation chains of the ferromagnetic clusters, it is expected to observe semiconducting or the semimetal type of conductivity described by variable range hopping or small polaron hopping regimes.

\begin{figure}
    \centering
    \includegraphics{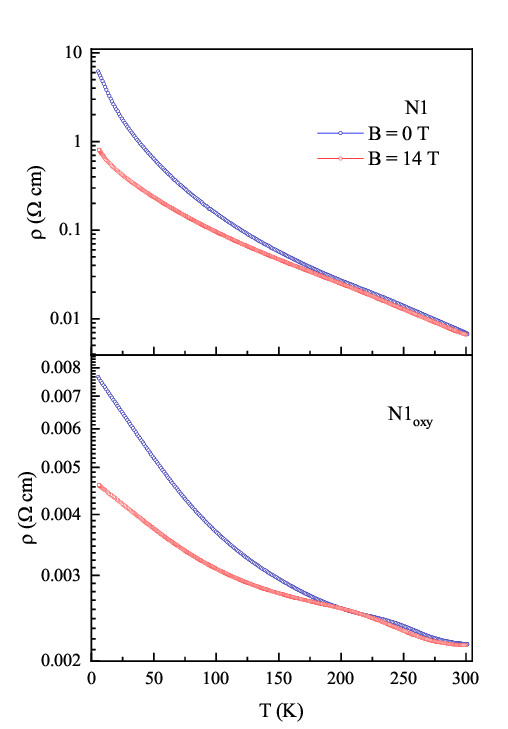}
    \caption{Temperature dependences of the resistivity for the as-prepared and oxidized Sr$_{0.9}$Nd$_{0.1}$CoO$_{3-\delta}$ samples measured at zero and an applied magnetic field of $B = 14$~T.}
    \label{fig:fig_R_T_N1}
\end{figure}

\subsection{Electrotransport properties}

\begin{figure}
    \centering
    \includegraphics{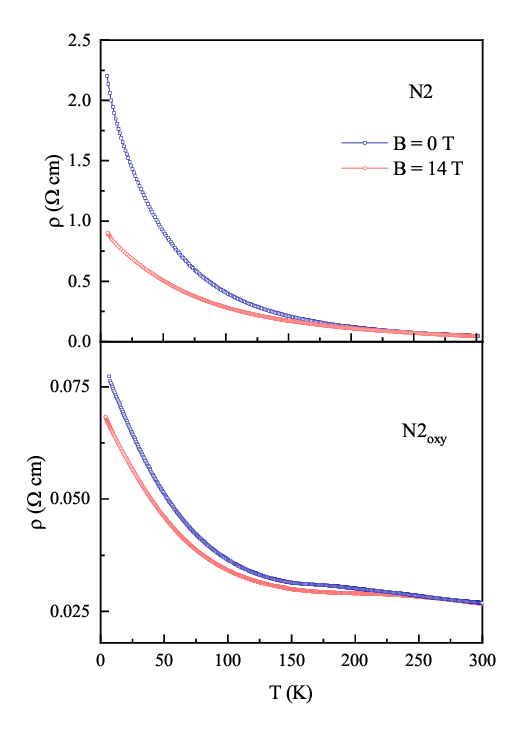}
    \caption{Temperature dependences of the resistivity for the as-prepared and oxidized Sr$_{0.8}$Nd$_{0.2}$CoO$_{3-\delta}$ samples measured at zero and an applied magnetic field of $B = 14$~T.}
    \label{fig:fig_R_T_N2}
\end{figure}

\begin{figure}
    \centering
    \includegraphics{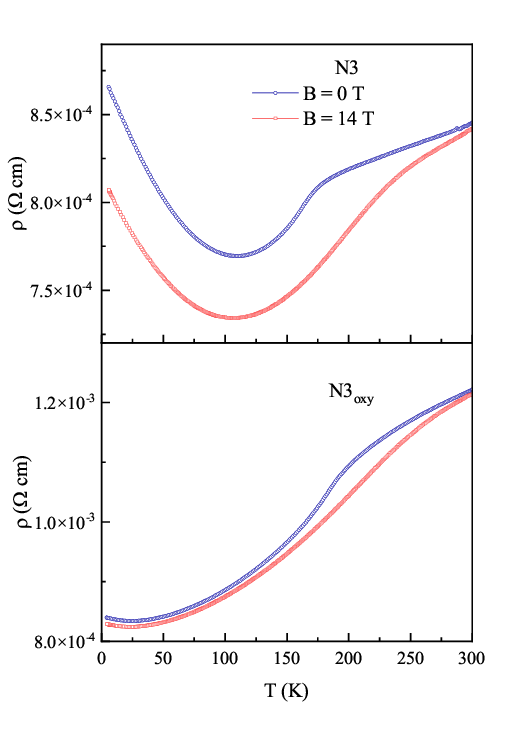}
    \caption{Temperature dependences of the resistivity for the as-prepared and oxidized Sr$_{0.7}$Nd$_{0.3}$CoO$_{3-\delta}$ samples measured at zero and an applied magnetic field $B = 14$~T.}
    \label{fig:fig_R_T_N3}
\end{figure}

Further studies of the electrotransport properties support the mixed spin-state scenario (ii) as the most probable model. Figures~\ref{fig:fig_R_T_N1}-\ref{fig:fig_R_T_N3} show the temperature dependences of the resistivity of the as-prepared and oxidized series measured in zero field and in an applied magnetic field of 14~T. For N1, having semiconductor-like type of conductivity, the oxygen index increase from 2.69 to 2.78 leads to a drastic decay in the resistivity values. Moreover, both N1 and N1$_\mathrm{oxy}$ possess a significant magnetoresistance at temperatures below 150~K. A very similar behavior is observed for samples N2 and N2$_\mathrm{oxy}$. This dramatic change in resistivity values distinguishes our results from those of Ref.~\cite{CHENNABASAPPA20206067}, where the near-stoichiometric oxidation case was considered.

A distinct behavior is observed for the  N3 sample, showing temperature-induced semiconductor to metal transition of the conductivity type upon heating. This is the only sample in the series that, without additional oxidation, has a sufficiently high oxygen index and up to 40\%  Co$^{4+}$. To understand the clear upturn of the resistivity curve at a temperature of $\sim$110~K  with the second anomaly at the metallic phase near the magnetic ordering temperature $T_\mathrm{mo}\sim$175~K, we should consider the difference between the N1, N2 and N3 compositions. 

In general, when Sr$^{2+}$ is substituted with Nd$^{3+}$, and the concentration of oxygen vacancies changes, the system can tolerate it in several ways: by a rearrangement of the Co--O bond lengths and angles \cite{Leighton_Lynn_PhysRevB.79.214420, Takami_2007}, which in extreme cases can lead to a complete structural transition \cite{CHENNABASAPPA20206067}, as in our case with the highly oxidized samples; or by a change in the spin state of some of the Co$^{3+}$ ions \cite{okimoto2021spin, Troyanchuk_pssb_2018,  Knizhek_2013_PhysRevB.88.224412, Lanovsky_pssb.202100636}. 

Due to the lack of NPD data for the samples with 10 and 30\% Nd, we cannot directly confirm how the bond angles in octahedral and anion-deficient layers change with increasing Nd$^{3+}$ content. Generally, there are two competing effects which can explain the observed properties: larger octahedral rotations and smaller Co--O--Co angles with increasing Nd$^{3+}$; and an improved Co--O network and  Co~3\textit{d}--O~2\textit{p} orbital overlap with oxygen uptake. Additionally, oxygen uptake promotes a higher percentage of Co$^{4+}$ ions and favors FM double-exchange. 

For N3, the temperature-induced semiconductor to metal transition of the resistivity can be caused by both the Co--O--Co bond angles strengthening, enhancing the Co~3\textit{d}--O~2\textit{p} orbital overlap, and temperature driven Co$^{3+}$  spin-state transition from mixed LS/HS to the new IS-containing mixed spin state.

Previously, a very similar temperature-driven crossover from a semiconducting spin-glass to metallic ferromagnetic state for La$_{1-x}$Sr$_{x}$Co$_{1-y}$Ni$_{y}$O$_{3-\gamma}$ was explained by changes in Co--O bond lengths and Co--O--Co bond angles within a single IS spin state of the Co$^{3+}$ ions \cite{Lanovsky_2024_PhysRevMaterials.8.114422}. Wu and Leighton \cite{Wu_Leighton_PhysRevB.67.174408} described the properties of the parent oxygen-stoichiometric La$_{1-x}$Sr$_{x}$CoO$_3$ within a model where both Co$^{3+}$ and Co$^{4+}$ are in the IS state. Related negative magnetoresistance was reported for metallic La$_{1-x}$Sr$_{x}$CoO$_3$ ($x=0.2, 0.3$) single crystals, where its magnitude near $T_\mathrm{C}$ approximately scaled with the square of the field-induced magnetization and was attributed to carrier–spin coupling within a double-exchange framework \cite{Yamaguchi1995_sing_cr}. In the present oxygen-deficient samples, the corresponding field response is additionally affected by oxygen-vacancy disorder, mixed Co valence and spin states, and the resulting inhomogeneous semiconducting transport.

With a further oxygen content increase, N3$_\mathrm{oxy}$ becomes metallic within the entire considered temperature range of 3--300~K.  It can be seen that generally the nature of the conductivity changes from semiconductor-like to metallic as the Nd$^{3+}$ and O$^{2-}$ contents increase, while both are interconnected. In addition, when moving from the AFM tetragonal layered $I4/mmm$ structure to the FM orthorhombic $Pmmm$ one, a decrease in the relative magnetoresistance values is observed. The oxidized N3$_\mathrm{oxy}$ compound is probably the only one in the whole series to which the itinerant electron model proposed by Bezdicka~et~al.~\cite{Bezdicka_SrCoO3_fully_st} should most likely be applied, along with the double-exchange and superexchange models.

\section{Conclusion}

In summary, the combination of high-resolution powder neutron diffraction  with synchrotron x-ray powder diffraction and complementary experimental techniques has revealed polymorphism in Sr$_{1-x}$Nd$_{x}$CoO$_{3-\delta}$ ($x = 0.1, 0.2, 0.3$) depending on the oxygen content. 

The as-prepared compounds adopt a layered $I4/mmm$ crystal structure with a $2a_p\times 2a_p \times 4a_p$ unit cell, exhibiting \textit{G}-type AFM magnetic ordering. Co$^{3+}$ ions within the oxygen-enriched layers are predominantly in the HS state, while the oxygen-deficient layers contain the majority of the Co$^{4+}$ ions. The bond angles within anion-deficient layers are strongly canted, which reduces the Co~3\textit{d}--O~2\textit{p} orbital overlap and weakens the potential strength of the FM Co$^{4+}$--O$^{2-}$--Co$^{3+}$ double-exchange interaction. Coupled with a half-filled oxygen position O2 within the $ab$ plane, this leads to spin-glass like magnetization dependences with a semiconducting-like resistivity. 

Partial oxidation suppresses the long-range layered ordering of oxygen vacancies of the $I4/mmm$  $2a_p\times2a_p\times4a_p$ structure, leading to a structural state with remaining preferentially occupied oxygen vacancies and octahedral tilting induced  by Nd$^{3+}$ substitution. A transition toward the $Pmmm$ space group with an $a_p\times a_p \times 2a_p$ unit cell and FM magnetic order occurs, featuring two of the Co--O--Co bond angles exactly at 180$^\circ$, while the other two are only insignificantly distorted.  The model assuming Co ions adopt the  LS/HS~Co$^{3+}$~+~Co$^{4+}$ state provides the best description of the observed magnetic and electrotransport properties, as well as the NPD results. The oxidized series is characterized by the presence of FM cluster-glass behavior and orders-of-magnitude lower resistivity and higher magnetization values compared to the tetragonal analogs. 

The magnetic properties can be understood in terms of the FM double-exchange operating within a magnetically inhomogeneous FM state, where residual oxygen vacancies and local structural distortions associated with the Nd$^{3+}$ substitution and mixed Co$^{3+}$/Co$^{4+}$ valence and spin states produce spatial variations of the exchange interactions.

Our results provide strong experimental support for the ideas proposed in Refs.~\cite{Marik_Toulemonde_oxygen, Fernandez_Toulemonde}, asserting that the driving force for the stabilization of the layered 314-type structure in cobaltites is the presence and long-range ordering of oxygen vacancies, rather than A-site cation ordering.

\begin{acknowledgments}

This research was financially supported by the Belarusian Republican Foundation for Fundamental Research Project No. F25ME-011. Part of this work was supported by the Ministry of Science and Higher Education of the Russian Federation within the framework of the State Assignment No. FZWM-2024-0011. The authors are grateful to Dr. Nikolay Kalanda and Dr. Alexander Petrov for their assistance in oxidizing the samples.

\end{acknowledgments}

\section*{DATA AVAILABILITY}
There are no publicly available research data or software supporting this manuscript. Requests for further information or data should be sent to the authors.

\appendix

\section{Additional crystal structural
parameters obtained from the refinement of high-resolution NPD data and refinement details}

\begin{figure}
    \centering
    \includegraphics{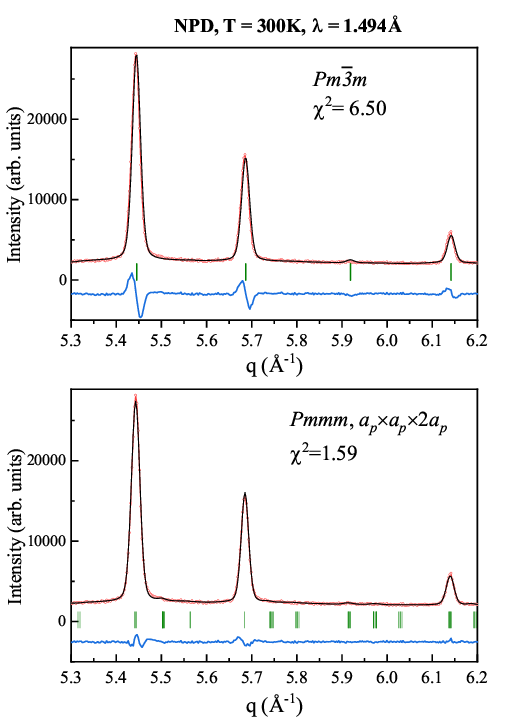}
    \caption{Enlarged regions of the room-temperature high-resolution NPD patterns of the N2$_{\mathrm{oxy}}$ sample, representing refinements using the cubic $Pm\overline{3}m$ (top panel) and orthorhombic $Pmmm$ (an $a_p \times a_p \times 2a_p$ unit cell, bottom panel) structural models. Both  refinements were performed using the Le Bail method to enable direct comparison.}
    \label{fig:appendix_fit_cube_orth}
\end{figure}

\textit{As-prepared series:} When studying layered cobaltites of the 314-type, an additional structural feature previously reported in the literature should be taken into account. An additional symmetry lowering from the tetragonal  $I4/mmm$ to the monoclinic $A2/m$ with two distinct supercells above and below the magnetic ordering temperature ($4\sqrt{2}a_p \times 4\sqrt{2}a_p \times 4a_p$ and $2\sqrt{2}a_p \times 4\sqrt{2}a_p \times 4a_p$, respectively) was observed for the Y-314 \cite{Lanovsky_pssb.202100636} and the Er-314 \cite{Ishiwata_PhysRevB.75.220406} cobaltites. 

This monoclinic distortion can be particularly detected in NPD patterns within the magnetically ordered phase, since it  produces additional Bragg reflections that cannot be described within the framework of tetragonal symmetry. Additionally, it may manifest itself  as a strongly broadened series of reflections associated with a monoclinically distorted phase and the absence of strict translational symmetry along the \textit{c} axis \cite{Lanovsky_pssb.202100636, Ishiwata_PhysRevB.75.220406, DUDNIKOV2020154629}. 

For the Nd-314 system considered in the present work, a slight broadening of the \textit{00l} series of reflections present at shorter wavelengths (1.494~\AA for neutrons and 1.25~\AA for synchrotron sources). All other Bragg reflections of the as-prepared samples can be perfectly described within the tetragonal $I4/mmm$ space group with a $2a_p \times 2a_p \times 4a_p$ unit cell down to a temperature of 1.5~K, both for the NPD and SXPD. The slight blurring of the peaks from the \textit{00l} series may be due to the random distribution of the ions in the A-position and statistical deviations in the distribution of oxygen ions along the \textit{c} axis, especially at the O2 8\textit{i}~(\textit{x},~0,~0) position in anion-deficient layers. No other indications of the symmetry lowering were observed. 

\textit{Oxidized series:} As previously noted in Sec.~III~A., a detailed examination of the powder diffraction patterns reveals a systematic splitting of a number of peaks, which is incompatible with a simple cubic space group. Figure~\ref{fig:appendix_fit_cube_orth} shows enlarged regions of the NPD patterns, collected at $T=300$~K and refined within the cubic $Pm\overline3m$  and orthorhombic $Pmmm$ $a_p \times a_p \times 2a_p$  structural models during the initial search in the Le Bail mode. The peak splitting is subtle for the available wavelengths, but sufficient to cause a significant difference in the description of the peak profiles in the proposed models. The cubic model produces systematic S-shaped residuals and cannot reproduce the asymmetric peak profiles observed for several reflections, whereas the orthorhombic model accurately enough reproduces these features. Le Bail refinements using $Pm\overline{3}m$ and $P4/mmm$ yielded $\chi^2$ values of 6.50 and 12.45, respectively, while Le Bail refinement within the orthorhombic $Pmmm$ model resulted in $\chi^2$~=~1.59, indicating that the observed diffraction pattern is intrinsically incompatible with the higher-symmetry descriptions. 

The main crystal structural parameters obtained from the NPD refinements are summarized in Table~\ref{tab:tab1_N2_npd}. 

\bibliography{Nd_Sr_Co_O}

\end{document}